\documentclass[preprint2]{aastex61}
\usepackage{multirow}
\usepackage{color}

\shorttitle{Gender and Questions at Cool Stars}
\shortauthors{Schmidt \& Douglas et al.}

\begin{document}

\title{The Role of Gender in Asking Questions at Cool Stars 18 and 19}
\author{Sarah J. Schmidt}
\affil{Leibniz-Institute for Astrophysics Potsdam (AIP), An der Sternwarte 16, 14482, Potsdam, Germany}
\email{sjschmidt@aip.de}
\author{Stephanie Douglas}
\affil{Department of Astronomy, Columbia University, MC 5246, 550 West 120th Street, New York, NY 10027}
\author{Natalie M. Gosnell}
\affil{Department of Physics, Colorado College, 14 East Cache La Poudre St., Colorado Springs, CO 80903}
\author{Philip S. Muirhead}
\affil{Department of Astronomy, Boston University, 725 Commonwealth Ave., Boston, MA 02215}
\author{Rachel S. Booth}
\affil{Astrophysics Research Centre, School of Mathematics \& Physics, Queen's University Belfast, University Road, Belfast BT7 1NN, UK}
\author{James R. A. Davenport}
\affil{Department of Physics \& Astronomy, Western Washington University, Bellingham, WA 9822}
\affil{NSF Astronomy and Astrophysics Postdoctoral Fellow}
\author{Gregory N. Mace}
\affil{McDonald Observatory \& The University of Texas, 2515 Speedway, Stop C1400, Austin, Texas 78712-1205}

\begin{abstract}
We examine the gender balance of the 18th and 19th meetings of the Cambridge Workshop on Cool Stellar Systems and the Sun (CS18 and CS19). The percent of female attendees at both meetings (31\% at CS18 and 37\% at CS19) was higher than the percent of women in the American Astronomical Society (25\%) and the International Astronomical Union (18\%). The representation of women in Cool Stars as SOC members, invited speakers, and contributed speakers was similar to or exceeded the percent of women attending the meetings. We requested that conference attendees assist in a project to collect data on the gender of astronomers asking questions after talks. Using this data, we found that men were over-represented (and women were under-represented) in the question sessions after each talk. Men asked 79\% of the questions at CS18 and 75\% of the questions at CS19, but were 69\% and 63\% of the attendees respectively. Contrary to findings from previous conferences, we did not find that the gender balance of questions was strongly affected by the session chair gender, the speaker gender, or the length of the question period. We also found that female and male speakers were asked a comparable number of questions after each talk. The contrast of these results from previous incarnations of the gender questions survey indicate that more data would be useful in understanding the factors that contribute to the gender balance of question askers. We include a preliminary set of recommendations based on this and other work on related topics, but also advocate for additional research on the demographics of conference participants. Additional data on the intersection of gender with race, seniority, sexual orientation, ability and other marginalized identities is necessary to fully address the role of gender in asking questions at conferences. 
\end{abstract}

\section{Introduction}
Science, technology, engineering, and mathematics (STEM) fields have long been dominated by the same group that dominates our culture at large -- primarily (white, straight, able, cis-) men. This is currently the case in Astronomy; The gender balance of the American Astronomical Society (AAS) was in 2013 reported to be 25\% women and 73\% men \citep{Anderson2014}, while the gender balance of the International Astronomical Union (IAU) is recorded to be (as of 1 March 2017) 17\% women and 83\% men.\footnote{\url{https://www.iau.org/administration/membership/individual/distribution/}, retrieved 2 March 2017} There is no one reason for the gender imbalance in Astronomy (and related fields), but some major factors include the cultural perception of brilliance as a necessary yet exclusively (white) male trait \citep{Leslie2015}, the reliance on tests like the general and physics GRE that have a documented gender imbalance \citep[as well as a racial imbalance;][]{Levesque2015,Miller2014} and the persistence of sexual harassment in the field.\footnote{for example \url{https://www.buzzfeed.com/azeenghorayshi/famous-astronomer-allegedly-sexually-harassed-students, accessed 14 March 2017}} 

There are a myriad of subtle effects that work in concert with those factors to devalue and de-emphasize the work of women. For example, faculty members reviewing resumes for lab managers that were identical except for the name consistently rated male applicants as more competent and deserving of more money (an average of \$4,000 more per year) than female applicants \citep{Moss-Racusin2012}. Papers by women are cited less, both by themselves \citep[i.e., self-citations;][]{King2016} and others \citep{Caplar2016}, and a review of Hubble Space Telescope and European Southern Observatory telescope proposal evaluations reveals that proposals with women as principal investigators are less likely to be highly ranked and receive valuable telescope time \citep{Reid2014,Patat2016}. 

While many of these studies focus on gender (and like our work, address gender as a binary that fails to acknowledge non-binary, agender, intersex, and gender fluid people), these factors are also persistent along other divisions such as race, ability/disability, sexual orientation, gender identity. People who are at the intersection of multiple marginalized identities (e.g., women of color, disabled women, disabled women of color, etc.) are usually subject to multiple barriers. For example, the study testing the hirability of lab managers \citep{Moss-Racusin2012} was based primarily on similar work \citep{Bertand2003} that found resumes with ``white'' names (Emily and Greg) were strongly favored compared to those with ``black'' names (Lakisha and Jamal). In practice, an applicant that is both female and black typically experiences biases for both identities. 

Academic conferences act as a microcosm of our field, magnifying some of the most pressing issues for inclusion in astronomy. Yet for many scientists, conferences are an extremely important opportunity to network and advertise both ourselves and our work, forming important collaborations and making connections essential to future employment. The participation of all scientists in conference settings is thus necessary for an inclusive field. 

The percent of women speakers is a statistic that has received significant attention, particularly in instances where it is lower than the overall fraction of women in the field, as was the case in these reports from American Microbiology and Australian Space Research meetings \citep{Casadevall2014,Horner2016}. An announcement of a Nueroscience conference without any women speakers prompted an initiative to track the gender of all speakers at Neuroscience conferences;\footnote{\url{https://biaswatchneuro.com/}} a similar initiative was undertaken by the Committee on the Status of Women in Astronomy (CSWA; \url{https://cswa.aas.org/percent.html}) between 2008 and 2013. Participation in a conference is not limited to giving a talk, however, and each conference interaction can be an important networking opportunity. 

After each scientific talk, a brief (2--10 minute) period is allotted for questions from the talk audience. The main motivation for this period is to foster a public scientific interaction that is not possible in other venues, benefiting the speaker, the questioner, and the audience. Multiple astronomers have noted that the question periods seemed to include more questions from men than warranted given their fractional attendance, and at the 2014 winter meeting of the American Astronomical Society (AAS 223), a web form was implemented to track the number of questions from men and women.\footnote{\url{https://github.com/jradavenport/Gender-in-Astro}} That initial study was described in a white paper \citep{Davenport2014b}, and a similar effort was undertaken at the 2014 annual National Astronomy Meeting \citep[NAM;][]{Pritchard2014}. Both found that the fraction of questions asked by women was lower than their attendance rate by about 10\%. This is concerning, as the visibility of asking questions in front of an audience is less available to women; additionally, sociological studies indicate that speaking more often and/or for longer in a group setting is linked to leadership and power \citep{Mast2002}.

Since the initial results from the Gender Questions project, data has been taken on a variety of meetings, including the last two meetings from the series of Cambridge Workshops on Cool Stellar Systems and the Sun (known as Cool Stars meetings). In this proceedings, we report on the gender balance at the last two Cool Stars meetings, including that of the organizers, speakers, and attendees, as well as the reported gender of people asking questions. In Section~\ref{sec:CSback}, we provide more detail on the Cool Stars meetings. The sources of our data are discussed in Section~\ref{sec:data}, in Sections~\ref{sec:questgender} and~\ref{sec:speakgender} we examine the gender of attendees who asked questions and the differences in the questions asked of female and male speakers. We provide brief resources and recommendations in Section~\ref{sec:conclusions}. A python notebook containing most of the analysis presented below can be found online at \url{https://github.com/jradavenport/Gender-in-Astro/blob/master/analysis/coolstars/douglas_coolstars_analysis.ipynb}.

\section{Cool Stars Overview}
\label{sec:CSback}
The first Cool Stars meeting was a small workshop organized at the Harvard Center for Astrophysics in Cambridge, MA in 1980, and the workshop series has continued for over 35 years. In 1993, the workshop series moved from Cambridge and started alternating between locations in Europe and the United States. Cool Stars meetings have occurred approximately every other year since then. In the past decade, Cool Stars meetings have grown to include over 300 participants from the US, Europe, and around the world. 

In this proceedings, we focus on the previous two Cool Stars meetings: Cool Stars 18 (CS18),\footnote{CS 18 website available at \url{http://www2.lowell.edu/workshops/coolstars18/index.html}} held 8-13 June 2014 in Flagstaff, AZ and Cool Stars 19 (CS19),\footnote{CS19 website available at \url{http://coolstars19.com/}} held 5-10 June 2016 in Uppsala, Sweden. Both conferences had a similar format, with a Sunday opening reception, five mornings of plenary talks, and three afternoons with parallel splinter sessions (split into three sessions at CS18 and into four sessions at CS19) running in the afternoon. Basic information for each Cool Stars meeting, including Scientific Organizing Committee (SOC) membership, attendees, and the speakers in each session, was available either online or upon request from the organizers. 

While these two conferences were similar, there were a few differences that may have affected gender balance of attendees and fraction of people asking questions. The location of the conference is likely a large factor; CS18 may have drawn a larger fraction of participants from the US, while CS19 drew a larger fraction of Europeans. The timing of the plenary sessions was different; CS18 featured slightly longer talks with shorter question periods, while CS19 focused on shorter talks with 5--10 minutes allowed after each one. Additionally, CS18 provided stand microphones for members of the audience with questions, while CS19 had volunteers run microphones to each person with a raised hand. The stand microphones may deter questions from women due to their public placement and the pressure of standing and walking to them; microphones with runners may mitigate that issue and are certainly an otherwise better option because mobility impaired people may have trouble accessing standing microphones.

\section{Data from Cool Stars}
\label{sec:data}
To address the role of gender in question/answer dynamics at CS18 and CS19, we gathered data both from conference materials and from a custom-designed webform that has been implemented at multiple conferences. 

\subsection{Conference Demographics}
\label{sec:demo}
The gender balances of organizers, attendees, and speakers (both invited and contributed) are essential to understanding both the conference as a whole and as a comparison for the fraction of women asking questions during talks. We obtained a list of SOC members and the list of speakers from the conference websites of each conference, and we obtained the final participant lists for each conference directly from the conference organizers.

In all cases, we assigned gender at first based on name. If the name wasn't known by our (US based) gender identifiers, a quick internet search was conducted to locate a picture of the attendee or, failing that, find a typical gender for people with that name. We recognize that this method is both imprecise and enforces a false gender binary on individuals who identify as non-binary, genderfluid, intersex, agender, and/or in any other way not reflected by the male/female binary. Many of our results are dependent on perception of gender, so this may reflect the biases of some of the conference attendees. 

\begin{figure}[t!]
\centerline{\includegraphics[width=\columnwidth]{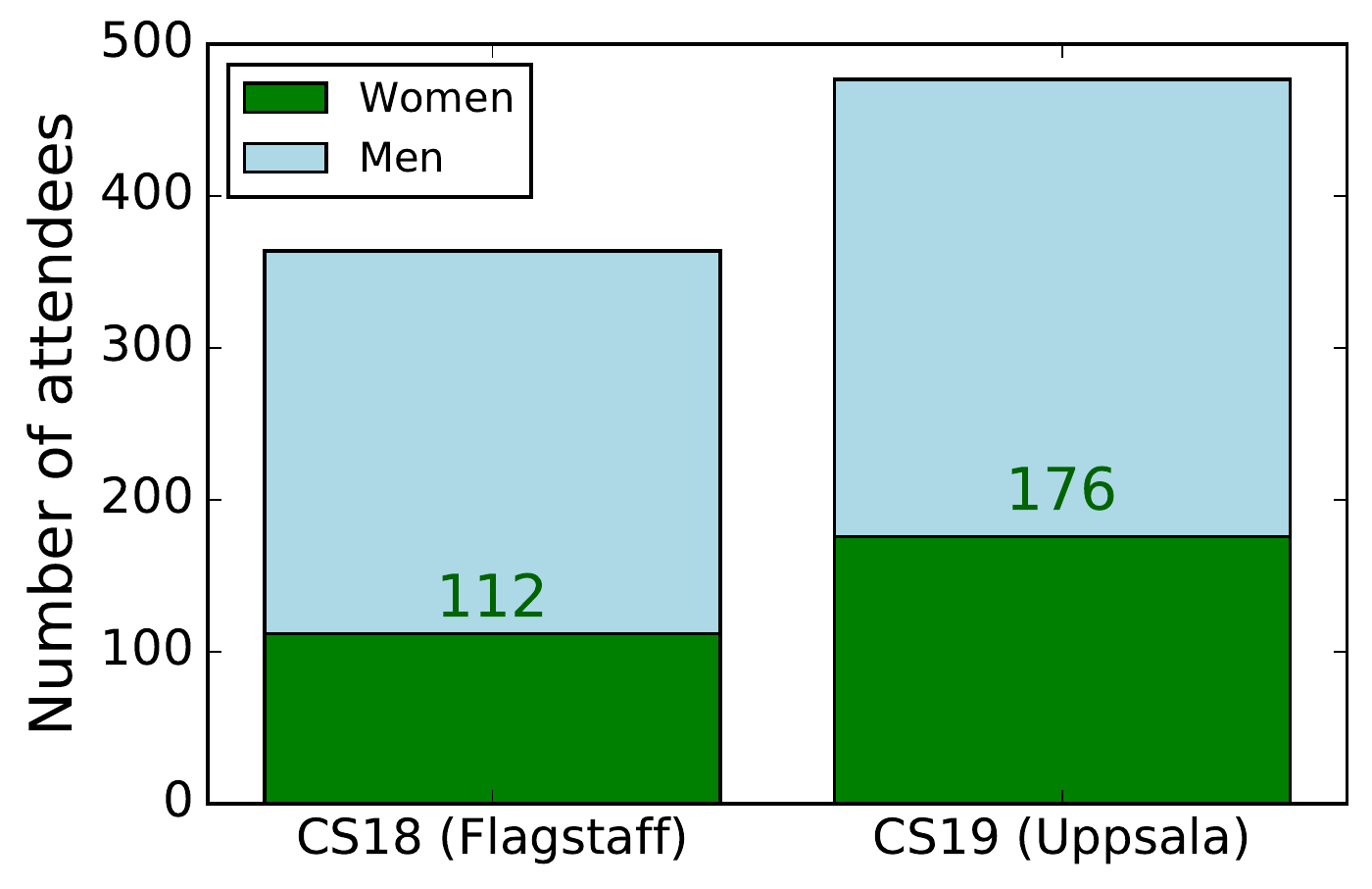}}
\centerline{\includegraphics[width=\columnwidth]{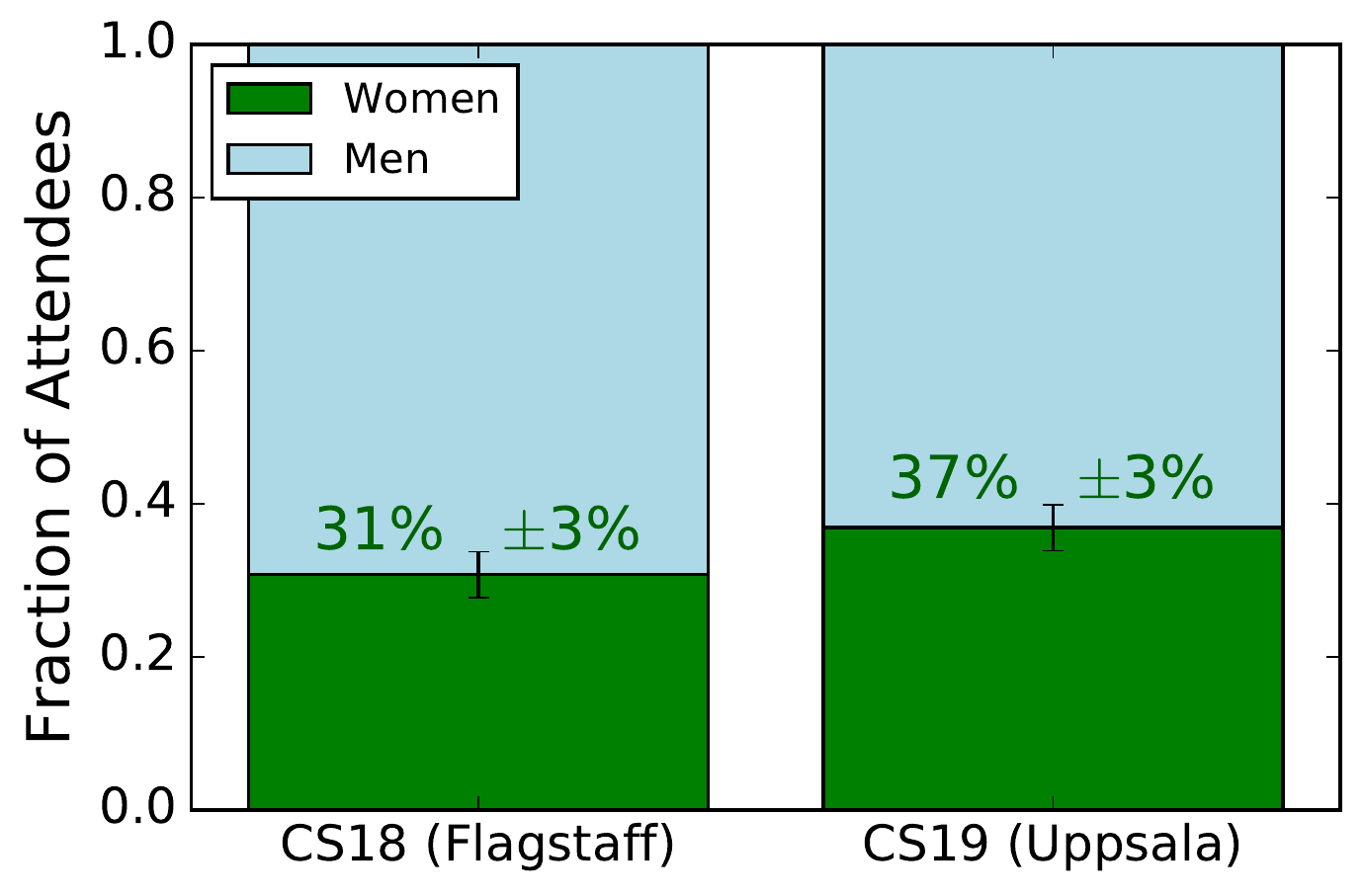}}
\caption{The total number of attendees (top panel) and fraction of female attendees (bottom panel) at CS18 and CS19. The uncertainties are assigned based on the ambiguities involved in assigning gender based on first names. The total number of attendees increased from CS18 to CS19, as did both the number of female attendees and the fraction of women. The data for this figure can be found in the first two rows of Table~\ref{tbl-1}.} 
\label{fig:attendees} 
\end{figure}

The results of our gender assignment are given in Table~\ref{tbl-1}. The total number of attendees rose from 364 to 447 between CS18 and CS19, and both the total number (112 to 176) and the percentage (31\% to 37\%) of women rose between the two years (as shown in Figure~\ref{fig:attendees}). The fraction of women attending both Cool Stars conferences is higher than the fraction of women astronomers; as reported above, women are 17\% of IAU members and 25\% of AAS members.

\begin{deluxetable}{clll} %google sheet with speaker data: https://docs.google.com/spreadsheets/d/1t53HvkQ54_eOzFhwWvOduBOfjz_j_Wl4WCgECmB0A-E/edit?usp=sharing
\tabletypesize{\scriptsize} \label{tbl-1}
\tablecaption{Conference Statistics} 
\tablewidth{0pt}
\tablehead{
\colhead{Group} & \colhead{Measurement} & \colhead{CS18} & \colhead{CS19}
}
\startdata
\multicolumn{4}{c}{Data obtained from conference materials (see Section~\ref{sec:demo})} \\
 \hline
\multirow{2}{*}{Attendees} & \# total & 364 & 477 \\
 & \% women\tablenotemark{a} & $31\pm3$~\% & $37\pm3$~\% \\
\hline
\multirow{3}{*}{SOC} & \# women & 7  &  10 \\ 
 & \# total & 20 & 25 \\
 & \% women & 35\% & 40\% \\
 \hline
\multirow{3}{*}{\parbox{2.5cm}{\centering Invited Plenary Speakers}} & \# women & 4 & 7 \\ 
 & \# total & 14 & 10 \\
 & \% women & 29\% & 70\% \\
\hline
\multirow{3}{*}{\parbox{2.5cm}{\centering All Plenary Speakers}} & \# women & 11 & 18 \\  
 & \# total & 29 & 38 \\
 & \% women & 38\% & 47\% \\
\hline
\multirow{3}{*}{Total Speakers} & \# women & 43 & 60 \\ 
 & \# total & 113 & 153 \\
 & \% women & 38\% & 39\% \\
\hline
\multicolumn{4}{c}{Data obtained from web form (see Section~\ref{sec:qdat})} \\
\hline
\multirow{3}{*}{\parbox{2.5cm}{\centering Speakers with Question Data}}& \# women & 29 & 34 \\ 
 & \# total & 76 & 83 \\
 & \% women\tablenotemark{b} & $38\pm11$~\% & $41\pm11$~\% \\
\hline
\multirow{3}{*}{Recorded Questions} & \# women & 46 & 67 \\ 
 & \# total & 223 & 264 \\ 
 & \% women\tablenotemark{b} & $21\pm5$~\% & $25\pm5$~\% \\
\hline
\multirow{3}{*}{\parbox{2.5cm}{\centering Talks With Question Data}} & \% All & 67\% & 54\% \\ 
 & \% Plenary & 90\% & 100\% \\
 & \% Splinter & 60\% & 37\% \\
\enddata
%\tablecomments{}
\tablenotetext{a}{Uncertainties from gender-ambiguous names.}
\tablenotetext{b}{Uncertainties from binomial confidence intervals.}
\end{deluxetable}

The difference between the field-wide statistics and the conference attendee fraction could reflect differences in women's representation between different subfields and at different levels of seniority. Statistics on gender and subfield are difficult to find, but \citet{Reid2014} note that during cycles 17--20 23\% of Hubble proposals in the Cool Stars category came from women, compared to 22\% overall. The combined fraction of female PIs with proposals in the cool stars, star formation, and exoplanet categories (which all participate in Cool Stars meetings to some extent) was 25\%. This could indicate a slightly larger fraction of women in the Cool Stars subfield.

Scientists who focus on Cool Stars (and those who attend the meetings) could also be disproportionately junior. Some of the areas covered by Cool Stars have been topics of study since well before the meetings began (e.g., Solar spectroscopy) but others have been developed more recently (e.g., brown dwarfs and exoplanets). This is notable because a newer field is typically a field with more women. The 2013 AAS demographics report notes that if members are divided into those born before and after 1980, the younger group is 40\% women and the older group is 21\% women. While age and seniority are not interchangeable, they are roughly correlated, indicating that a higher fraction of junior astronomers are women while a smaller fraction of senior astronomers are women. 

We also gathered data on the total number and percent of women among the conference Scientific Organizing Committee (SOC), the invited speakers, and all speakers. This data was gathered using the conference webpages, and gender was assigned using the same method as for the attendee list. 

The SOC grew from a 20 person committee planning CS18 to a 25 person committee for CS19. Both the number and fraction of women on the SOC rose between the two conferences, from 7 (35\%) to 10 (40\%). This rise is similar to the rise in the fraction of women attendees. The invited plenary speakers for CS18 were 29\% women, while the invited plenary speakers for CS19 were 70\% women. Both the SOC and the invited speakers are advertised during the registration period for the conference, and as such they can have an effect on the composition of conference attendees. This is explicitly used to select the scientific focus of a conference, and is likely to also impact the conference demographics. While the increase in women between CS18 and CS19 is the result of several factors, the high fraction of female SOC members and invited plenary speakers is likely to have played a role. 

The total number of speakers increased from 113 to 153 between CS18 and CS19, primarily as a result of the change from three to four concurrent afternoon parallel sessions. Overall, the fraction of women speakers was remarkably constant, rising from 38\% to 39\% between the two conferences. In CS18, women were over-represented as speakers compared to conference attendees (38\% compared to 31\%) while in CS19, the fraction of women speakers was more representative of the conference demographics as a whole (39\% compared to 37\%). Because Cool Stars conference are a series (rather than conferences in isolation), it is possible that the high fraction of women speakers at CS18 encouraged a larger fraction of women to attend CS19. 

\subsection{Gender of Questioners}
\label{sec:qdat}
\begin{figure}[t!]
\vspace{12pt}
\centerline{\includegraphics[width=0.9\columnwidth]{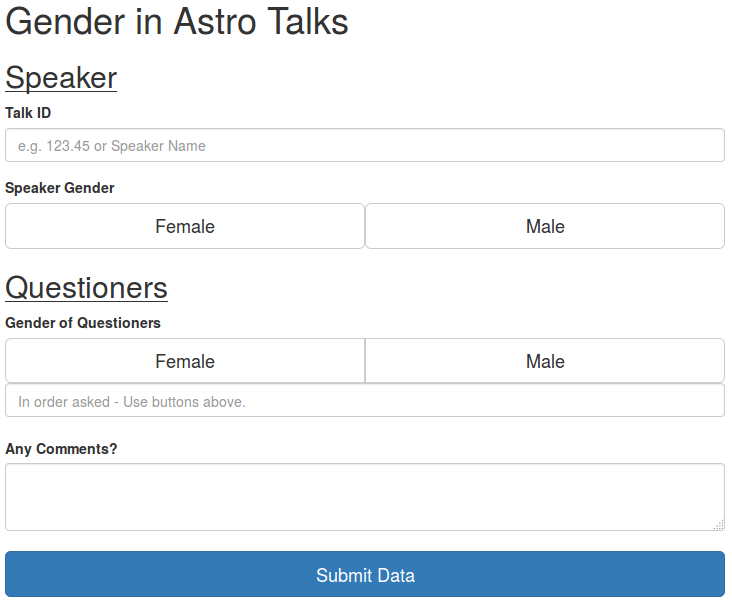}}
\vspace{12pt}
\centerline{\includegraphics[width=0.9\columnwidth]{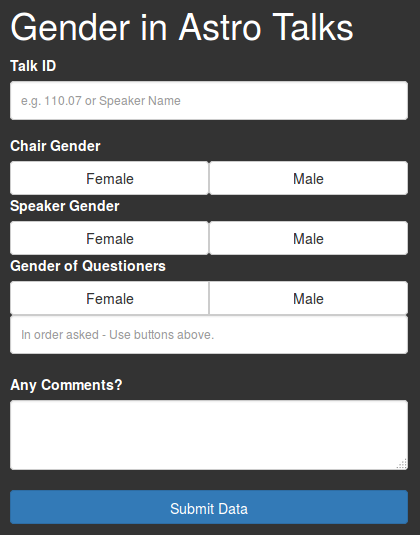}}
\caption{The web form used for CS18 (top) and for CS19 (bottom) to gather data on the gender of the speaker, chair, and questioners. Both forms include a free entry space for Talk ID, ``female'' and ``male'' buttons for Speaker Gender, and ``female'' and ``male'' buttons for Speaker Gender, and a free entry box for comments. The buttons for Speaker Gender generate a line of text that fills in a text box. The form used for CS18 does not include gender buttons for the session chair, while the CS19 form does.} 
\label{fig:form} 
\end{figure}

Participants at CS18 and CS19 were given (via twitter) an opportunity to fill out a web form first identifying the talk by inputting the speaker name, talk number, and/or session, then selecting male or female for the speaker, the session chair (at CS19 - not yet implemented at CS18), and for each person asking questions. The form (shown in Figure~\ref{fig:form}) was designed to be equally accessible from smartphones, tablets, and computers. The majority of volunteers accessed the form during the question and answer portion of the talk, tapping a button as each question was asked. Each question was represented by a ``M" or ``F" in the data; a talk with three questions asked by a man, a woman, and then a man would result in a string of ``MFM." Data submitted through the webform were automatically added to a csv file, which was cleaned by hand. 

At Cool Stars 18, 223 questions for 76 talks were reported, divided into 87 questions for 26 plenary talks and 136 questions for 50 talks in six splinter sessions. This includes 90\% of the total plenary talks, but only 54\% of the splinter talks. At Cool Stars 19, 264 questions for 83 talks were reported. These were divided into 151 questions for 40 plenary talks and 113 questions for 43 talks in five splinter sessions. This included all plenary talks, but only 37\% of splinter talks. Thirty-seven talks at Cool Stars 18 and thirty-six talks at Cool Stars 19 were reported more than once; conflicting reports are noted in the table, which can be found at \url{https://github.com/jradavenport/Gender-in-Astro/tree/master/data/cs18} and \url{https://github.com/jradavenport/Gender-in-Astro/tree/master/data/cs19}.

Overall, women asked 21\% of the questions at CS18 and 25\% of the questions at CS19. The increase between the two conferences is not statistically significant (as the confidence intervals are 5\%). If the rise is a real effect, it may simply be a reflection of the increasing percentage of women attendees (31\% at CS18, and 37\% at CS19). At both conferences, men asked more questions (and women asked fewer questions) than would be expected from their attendance rate. 

\begin{figure}[!t]
\centerline{\includegraphics[width=\columnwidth]{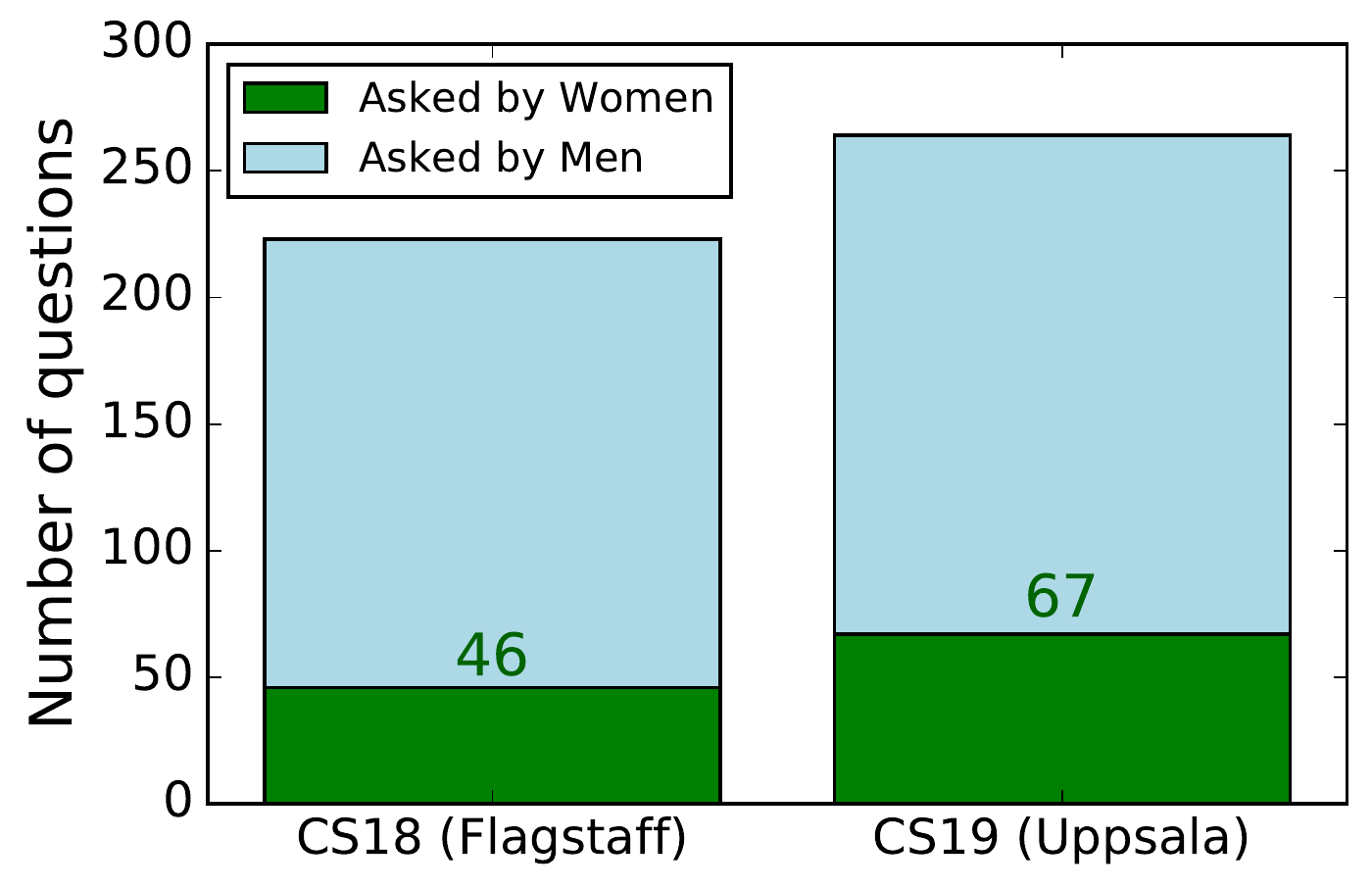}}
\centerline{\includegraphics[width=\columnwidth]{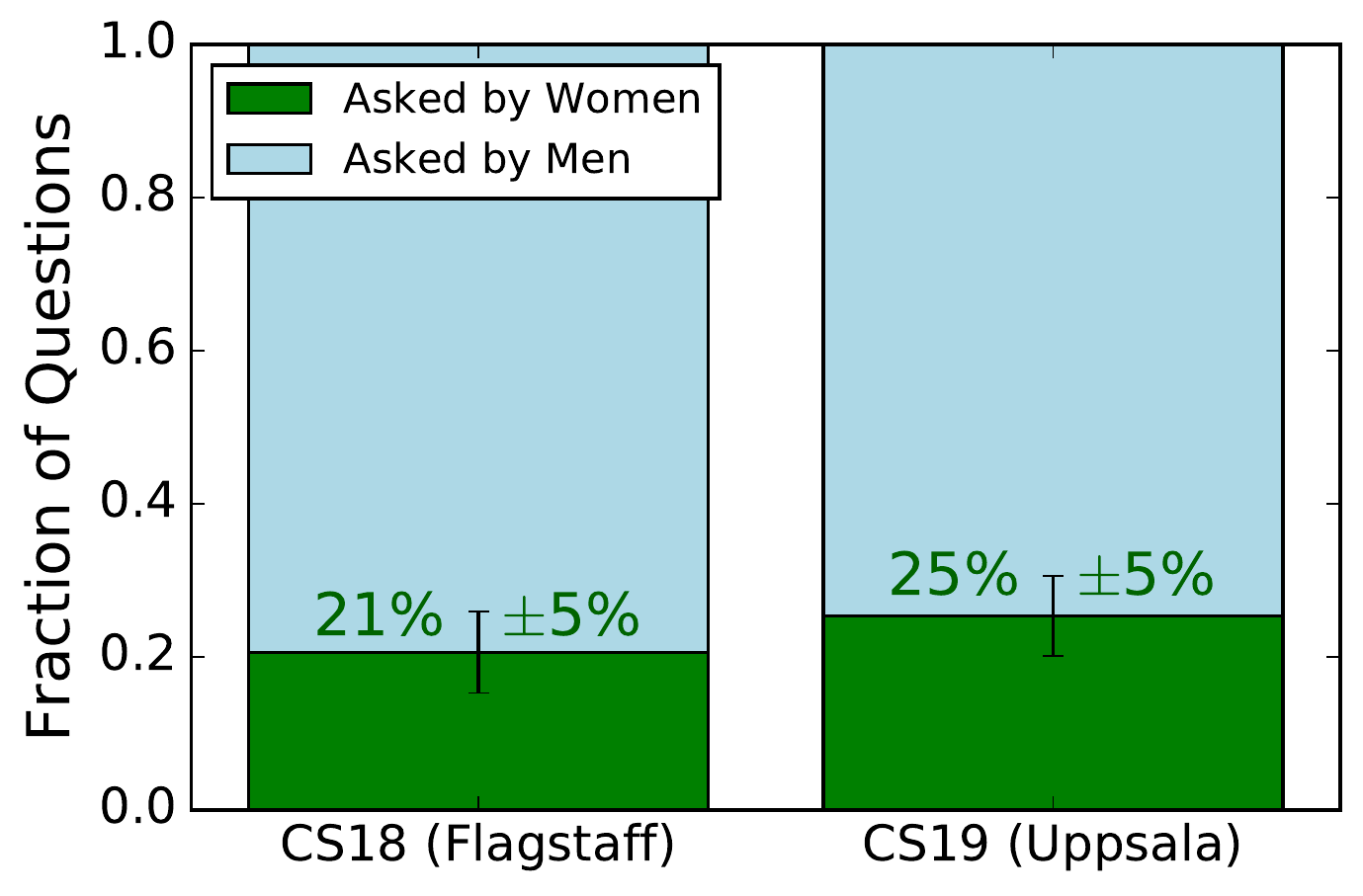}}
\caption{The total number of questions (top panel) and fraction of questions (bottom panel) asked by men and women recorded during CS18 and CS19. The overall number of questions increased from CS18 to CS19, as did both the number and fraction of questions asked by women. The 5\% uncertainties are based on Poisson counting errors. The total fraction of questions asked by women is low compared to the percent of women attending the conference (21\% compared to 31\% and 25\% compared to 37\%). The data shown here can be found in Table~\ref{tbl-1}.} 
\label{fig:raw_qs} 
\end{figure} 

One of the main possibilities suggested for this effect is seniority: the gender balance of senior (faculty and equivalent) astronomers is notably different from that of junior (student and postdoctoral) astronomers. While there is no data on the question balance between junior and senior astronomers, many feel that that senior astronomers typically ask questions after talks. The percent of questions from women matches more closely to the AAS membership on senior levels (21\% women) than overall (30\% women) or on junior levels (40\% women). 

Seniority is not the only factor at play, however. Junior astronomers do ask questions, and if they are less well known, they may be less likely to be remembered than the senior astronomers asking. Additionally, gender and seniority cannot be considered as two independent factors. Research on the joint roles of seniority and gender in volubility \citep{Brescoll2011} has found that men with more power/seniority speak more often than junior men, but women with more power/seniority limit themselves to speak only as often as more junior women, primarily due to fears that they will be perceived negatively for speaking too much. As part of their study, \citet{Brescoll2011} also found those fears to be justified, as women are frequently penalized for talking too much. 

\section{Examining Questioner Gender}
\label{sec:questgender}
There are multiple factors that may work to encourage (or discourage) questions from women (or men). A close examination of those factors may reveal certain environments that foster a more inclusive question and answer session. Here, we focus first on the genders of the speaker and the session chair and then on the number of questions in each post-talk question period. 

\subsection{Speaker and Chair Gender}
\begin{figure}[!t]
\centerline{\includegraphics[width=\columnwidth]{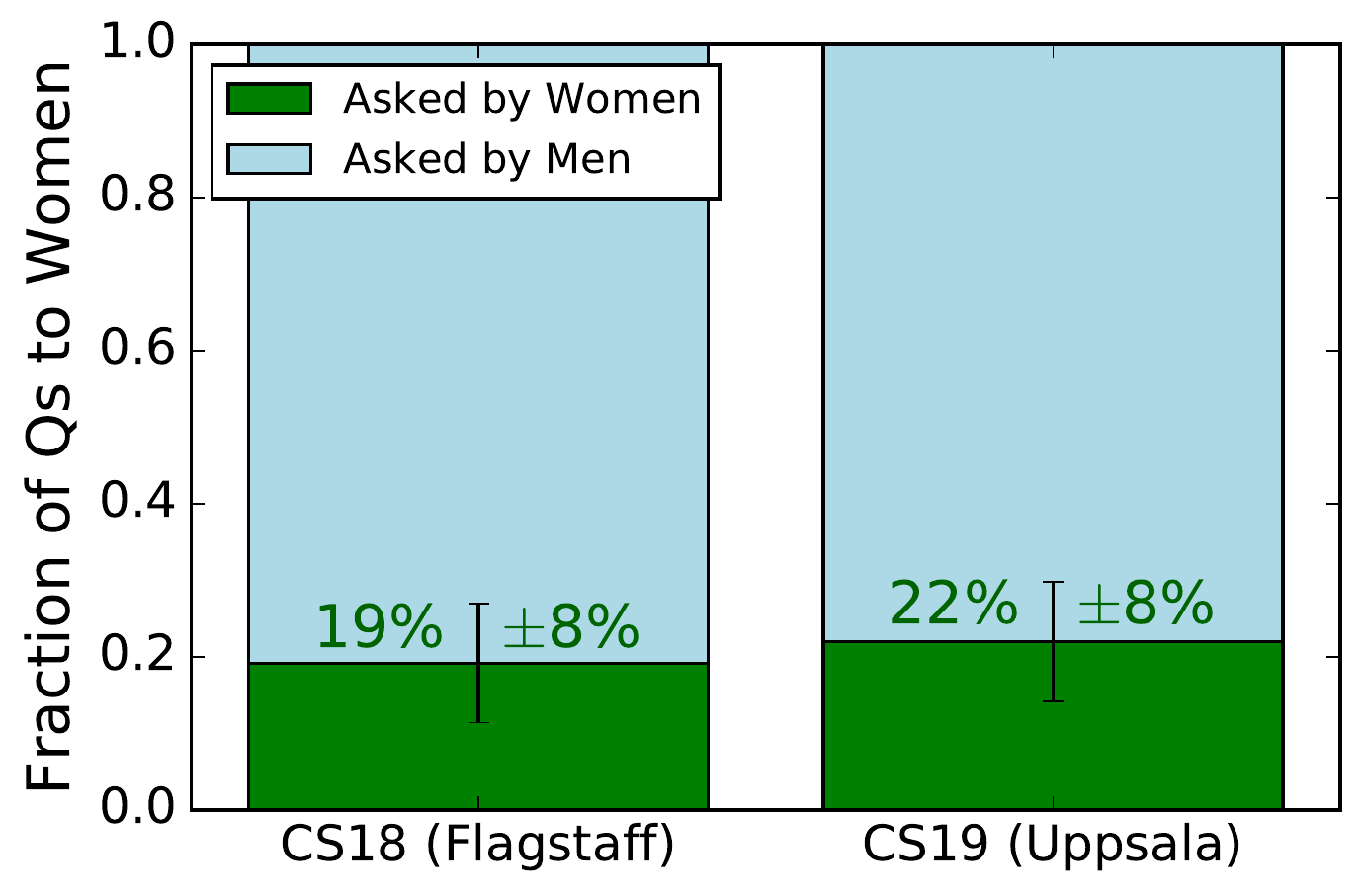}}
\centerline{\includegraphics[width=\columnwidth]{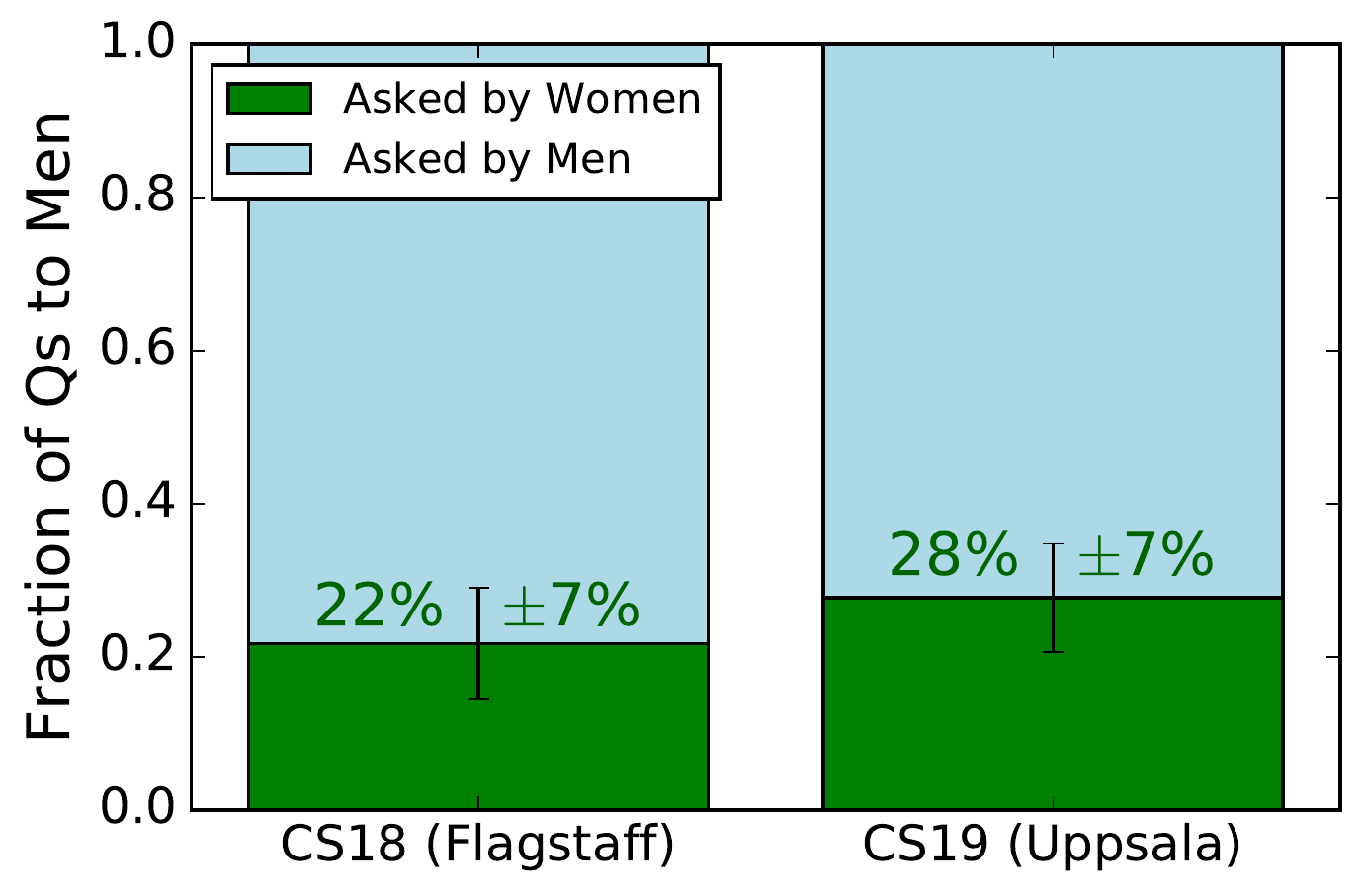}}
\caption{The fraction of questions by women addressed to female speakers (top panel) and to male speakers (bottom panel) recorded during CS18 and CS19. At CS18, women asked 19$\pm$8\% of the questions to women and 22$\pm$7\% of the questions to men. At CS19, women asked 22$\pm$8\% of the questions to women and 28$\pm$7\% of the questions to men. Overall, female speakers were asked more questions by men than male speakers.} 
\label{fig:qs_to_gender} 
\end{figure}

Previous results indicate that the gender of both the speaker and the chair of the session may have some effect on the number of questions from women and men \citep{Davenport2014b,Pritchard2014}. In Figure~\ref{fig:qs_to_gender}, we show the fraction of women and men who asked questions of male and female speakers. At CS18, the gender of questioners was similar; 19\% of questions to female speakers and 22\% of questions to male speakers came from women. At CS19 these fractions differed: 22\% of questions to female speakers and 28\% of questions to male speakers came from women. In each case, female speakers had fewer questions from women than male speakers. This indicates that at Cool Stars meetings, women are not necessarily more comfortable asking questions of women - or if they are, another factor mediates that effect. One suggestion is that some male audience members feel more strongly they need to provide comments or suggestions for the work of women that they feel they don't need for men. These and other reasons cannot be evaluated without additional evidence. 

\begin{figure}[t!]
\centerline{\includegraphics[width=\columnwidth]{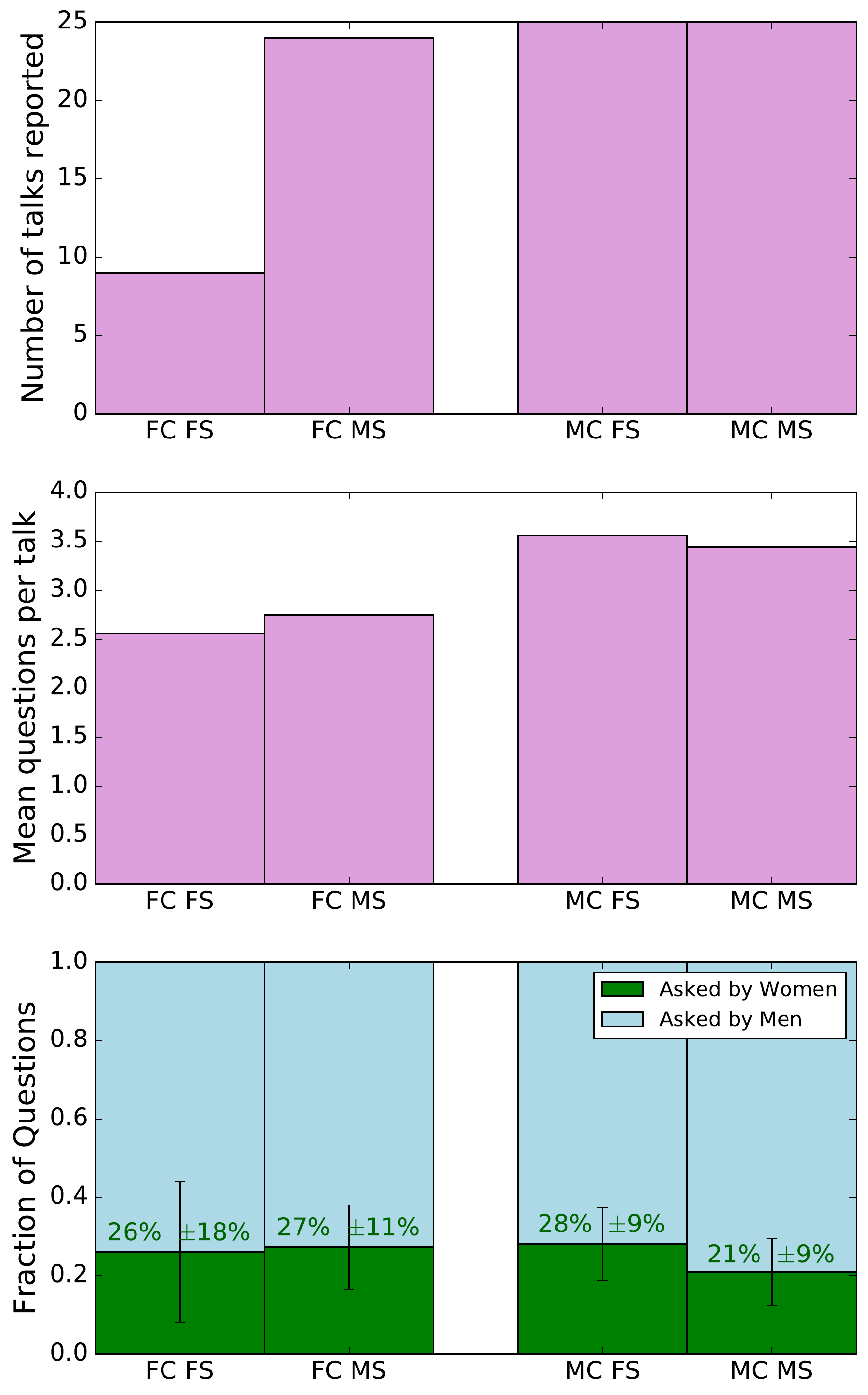}}
\caption{The number of talks (top), mean number questions per talk (middle), and fraction of questions asked by women after each talk (bottom), for talks divided into categories based on the gender of the chair (FC = female chair, MC = male chair) and the gender of the speaker (FS = female speaker, MS = male speaker). Data displayed in this Figure can be found in Table~\ref{tbl-2}.} 
\label{fig:chair_dist} 
\end{figure}

No data on session chair gender was taken during CS18, but during CS19 the gender of the chair was incorporated into the webform. We compare the number of talks, mean questions per talk, and fraction of questions from women based on speaker and chair gender, shown in Figure~\ref{fig:chair_dist}. The total number of talks with female chairs (FC) and male speakers (MS) was the same as those with male chairs (MC) and either gender of speaker, but the number of talks with both female chairs and female speakers was low. The male chairs allowed a larger number of questions per talk for speakers of any gender - 3.5 compared to 2.5. Both of these results are explained in part by the plenary sessions; they had a high fraction of female speakers and long question periods, but were all chaired by men. 

\begin{deluxetable}{l|ll|ll} 
\tabletypesize{\scriptsize} \label{tbl-2}
\tablecaption{Talk Statistics by Chair and Speaker Gender} 
\tablewidth{0pt}
\tablehead{ \colhead{Chair Gender} & \multicolumn{2}{c}{Female}  & \multicolumn{2}{c}{Male}  \\
\colhead{Speaker Gender} & \colhead{Female} & \colhead{Male} & \colhead{Female} & \colhead{Male}
}
\startdata
\# of talks reported & 9 & 24 & 25 & 25 \\
Mean questions per talk & 2.6 & 2.8 & 3.6 & 3.4 \\
Questions from women & 26$\pm$18\% & 27$\pm$11\% & 28$\pm$9\% & 21$\pm$9\% \\
\hline
\enddata
%\tablecomments{}
\end{deluxetable}

We find that the gender of the session chair has minimal impact on the fraction of questions asked by women. Women asked $27\% \pm 9\%$ of questions in sessions chaired by a woman, while they asked $25\% \pm 6\%$ of questions in sessions chaired by a man. The combinations of female chair and either female of male speaker yielded similar results (26\% vs. 27\% respectively) but the combination of male chair and female speaker yielded more questions from women (28\%) than the combination of male chair and male speaker (21\%). This difference is not significant compared to the uncertainties, however, nor is it as strong as the differences found in previous work. At AAS 223, \citet{Davenport2014b} found women asked 34\% of the questions in sessions with a female chair and 20\% of the questions in sessions with a male chair. The same effect was found at NAM \citep{Pritchard2014}, where women asked 22\% of the questions in sessions chaired by women and 16\% of the questions in sessions chaired by men. 

It's unclear why chair gender would have less impact at Cool Stars when the discrepancy was pronounced at both AAS 223 and NAM. Each of these meetings was a different size, and the community in attendance could be expected to have a different set of relationships. Because Cool Stars is restricted to a single subfield, it is perhaps more likely that the attendees are already familiar with the topic, the speaker, and the chair - a factor that could mitigate gender differences. The splinter sessions were also organized by small committees that typically included at least one woman; that could mitigate the effect of a male chair on questions from women. Gathering data from a wide variety of meetings may provide additional insights into this discrepancy. 

\subsection{Total Question Number}
\begin{figure}[t]
\centerline{\includegraphics[width=\columnwidth]{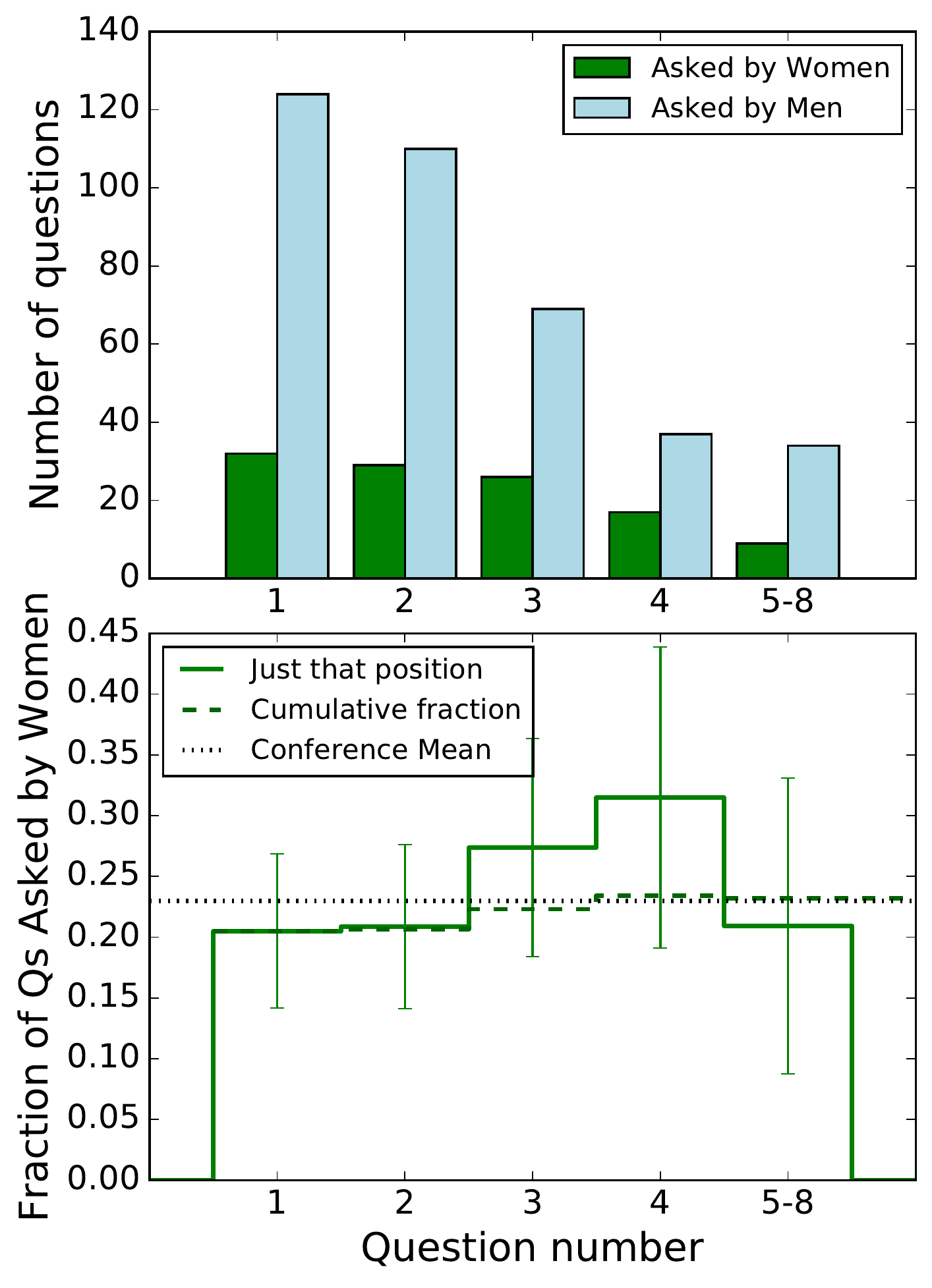}}
\caption{The number (top panel) and fraction (bottom panel) of questions asked by women and men at both Cool Stars meetings as a function of question order. The total number of questions decreases with question order from $\sim$150 for the first question to $\sim$40 for the fifth through eight question. The fraction of questions asked by women is is below total mean for the first and second questions, then increases for the third and fourth questions. These variations (5--10\%) are not significant compared to the Poisson uncertainties (10--20\%).} 
\label{fig:orderboth} 
\end{figure}

The total number of questions allowed after each talk could also be a mitigating factor in the total gender balance of question askers. During a longer question period, women who are shy or unsure of their questions may take time to gain confidence, or may realize their question is just as relevant as the others being asked. At the NAM meeting, there was a strong trend of an increasing fraction of questions from women as the total number of questions increased \citep[see their Figure~A;][]{Pritchard2014}. To test for a similar result at the Cool Stars meetings, we examined both the number and fraction of questions asked by women and men as a function of the total questions recorded, shown in Figure~\ref{fig:orderboth}. We combined data from both meetings to decrease the associated uncertainty due to small numbers. 

The mean fraction of questions asked by women for CS19 and CS19 combined was 23\%, but the fraction of first questions asked by women was 21\%. The fraction of second questions asked by women was similar (21\%), but the fraction of third questions asked by women was 28\%. The cumulative fraction of the first to third questions asked by women still fell slightly below the mean at 22\%. The cumulative fraction of the fourth and fifth to eighth questions asked by women was 23\%. This increase in the fraction of questions asked by women mirrors the trend found by \citet{Pritchard2014}, but is not significant compared to the uncertainties. Further data is needed to find if this trend is specific to conference size, type, or subfield. 

\section{Questions for Male or Female Speakers}
\label{sec:speakgender}
\begin{figure}[t]
\centerline{\includegraphics[width=\columnwidth]{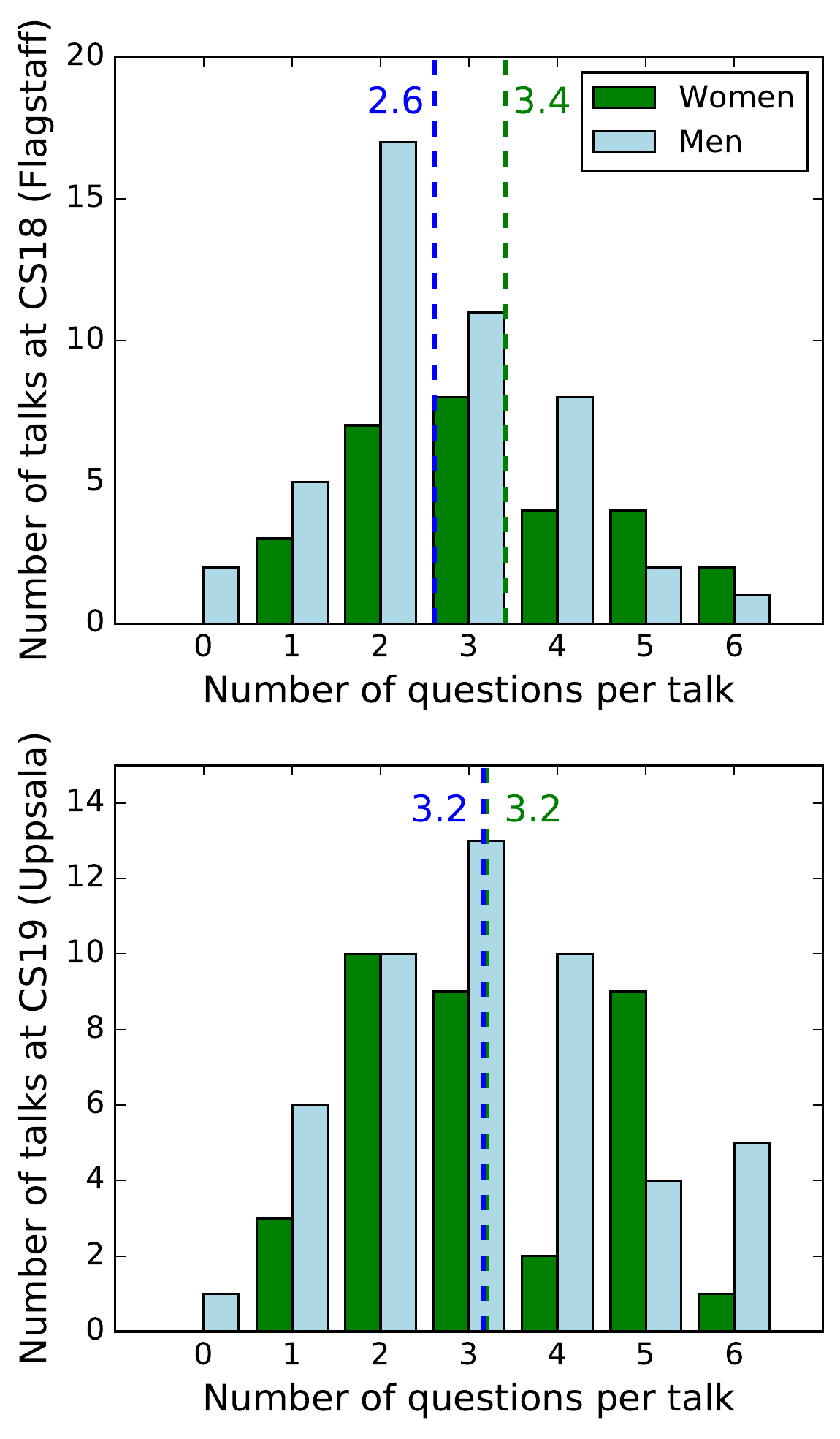}}
\caption{The number of talks where a specific number of questions were asked at CS18 (top) and CS19 (bottom). The data for male (blue) and female (green) speakers are separated, and the mean number of questions per talk for each group is also shown (dashed line). At CS18, men were asked on average one less question then women (2.6 compared to 3.4 questions per talk), while as CS19 men and women were asked an equal number of questions (3.2 questions per talk).} 
\label{fig:speaker_gender} 
\end{figure}

In addition to quantifying the number and percentage of questions asked by women, we can also use the question data to examine the different numbers and percentages of questions asked the male and female speakers. Figure~\ref{fig:speaker_gender} shows the distribution in the number of questions asked to male and female speakers at both CS18 and CS19. At CS18, men were asked an average of 2.6 questions compared to 3.4 asked to women. Though that is nearly one more question per talk, the difference is not significant compared to the Poisson uncertainties on the fraction. At CS19, men and women were both asked 3.2 questions on average. 

While there is no significant difference in the number of questions asked men and women, there are a variety of competing factors that could influence these numbers. A large number of questions could be the result of an interesting, stimulating talk that is generally admired, or a similarly excellent talk that inspires ``grilling'' from the community. A small number of questions could indicate an unusually clear talk, a talk that was difficult for the majority of the audience to understand, or simply a short question period. Further classification of the types of questions asked could clarify some of these differences. 

\section{Conclusions}
\label{sec:conclusions}
The last two Cool Stars meetings, CS18 and CS19, included a fraction of female attendees (31\% and 37\%) larger than that of the astronomical field as a whole (27\% from the AAS, 17\% from the IAU). This could be in part due to the contributions from more junior astronomers, a demographic that has a fraction of women closer to 40\%. It is also likely to depend on the fraction of female SOC members (35\% and 40\%) and invited plenary speakers (38\% and 70\%), as those are listed on the website before registration is completed. The total fraction of female speakers (38\% and 39\%) was higher than the fraction of female attendees at both meetings, though at CS19 the fractions were closer to equal. 

Despite an relatively high overall fraction of women involved in the two meetings, the fraction of questions asked by women was low (21\% and 25\%) compared to the fraction of female attendees. The difference between the fraction of women attending the conference and the fraction of questions asked by women (31\% compared to 21\% for CS18; 37\% compared to 25\% for CS19) is in line with the findings from previous results from the AAS and NAM meetings \citep{Davenport2014b,Pritchard2014}.

The results from the AAS and NAM meetings show strong dependencies of the fraction of questions asked by women on the gender of the speaker and chair as well as on the total number of questions asked. The data from the Cool Stars meetings do not reproduce these differences, indicating that they may not be consistent across different meetings sizes and/or subfields. We also examined the number of questions asked to male and female speakers, finding that they were nearly the same at CS18 and exactly the same at CS19. 

\subsection{Recommendations}
Due to the lack of correlations between the fraction of questions asked by women and any other examined variable, we have no recommendations based solely on the data for the Cool Stars meetings. As part of this Cool Stars hack day project, we compiled a list of recommendations based on the results of \citet{Davenport2014b} and \citet{Pritchard2014}, as well as from other sources. A static version of this document is contained in Appendix~\ref{sec:prac} and a live editable document is available at \url{https://github.com/jradavenport/Gender-in-Astro/wiki/Draft-of-Best-Practices}.

We also recommend that further data be collected on the gender questions project. This is not only to work towards a better understanding of the trends and patterns and their differences across many conferences, but also because the act of collecting data and advertising the project raises awareness that there are strong differences in the participation of men and women in the public forum of a question and answer session. 

And last, we recommend that the community works to encourage and respect different forms of interaction in addition to the public question forum. Often, the twitter feed of a conference contains not only summaries of the conference talks, but also an array of questions that can be answered in real time by our fellow astronomy tweeps. We should aim to work in a scientific community where each scientist can interact and be visible in the ways that are more comfortable and familiar than the public forum of a conference stage. 

\subsection{Future Work}
There are (at least) two ways the current project fails to reflect the astronomical community. The current incarnation of the survey reinforces a false gender binary (through the ``M'' and ``F'' buttons), further marginalizing people who identify as non-binary, genderfluid, intersex, agender, and/or in any other way not covered by those two labels. A third button was considered, but was not implemented because it is unclear that it would be appropriately used by those collecting data. If the questioner has not self-identified to the person recording the data, there would likely be a hesitancy to identify someone as non-binary, with a high chance of misgendering that person. Another strategy could be to use a similar strategy as the ``Are Men Talking Too Much'' site\footnote{\url{http://arementalkingtoomuch.com/}; \url{https://github.com/cathydeng/are-men-talking-too-much}}, which uses ``dude'' and ``not a dude'' as the gender classifiers. Input on these ideas is welcomed by the authors and can be implemented in future iterations of the survey.

The current incarnation of a survey also fails to recognize intersectionality. Men, women, and non-binary people are not monolithic groups that have the same axes of privilege and oppression, but are instead made up of a collection of people that can be further divided on race/ethnicity, disability, sexual orientation, economic background, culture, national origin, age, and seniority. Each of these factors determine not only whether a person is more or less likely to ask a question, but also whether they may be attending the conference in the first place. The proper way to address each classification is to request that conference participants self-identify, then track questions by name, later matching to demographic data. That undertaking would likely require the support of trained sociologists.

\begin{figure}[t!]
\vspace{12pt}
\centerline{\includegraphics[width=0.9\columnwidth]{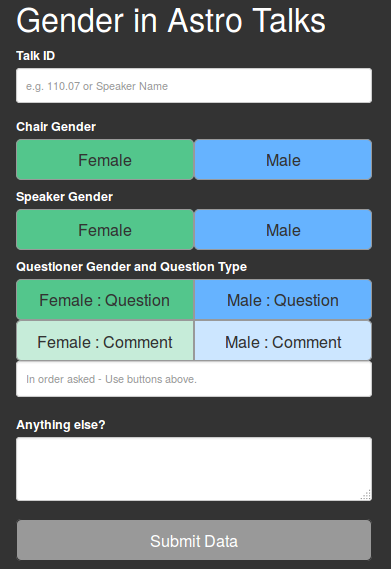}}
\caption{Current implementation of the web form for the ongoing gender in questions project. The majority of the form is the same as those in Figure~\ref{fig:form}, but buttons are added to capture feedback on whether the audience response is in the form of a question or a comment.} 
\label{fig:form_new} 
\end{figure} 

A combination of form responses (in the free response box) and feedback from conference attendees indicated that some audience members would make a comment instead of a question. Some additionally noted a pattern of men commenting on the work of women, while male speakers were instead asked questions. If this trend bears out, it could represent a devaluing of the work presented by female speakers i.e., men are trusted to be the experts in their topic, while women are given ``advice'' or ``tips'' by a commenter who believes they are the expert. To gather data on this phenomenon, we have implemented an updated web form that has separate buttons for ``question'' and ``comment'' (shown in Figure~\ref{fig:form_new}). The webform was implemented at the January 2016 AAS meeting (AAS 229) but results have not yet been examined. 

\acknowledgments
The authors wish to thank the organizers and sponsors of the Cool Stars 19 Hackathon \& Collaborative Writing Day. 
Thanks to Gerard Van Belle for providing data from CS18, including a list of attendees. 
We have made use of the {\it Astropy} package \citep{astropy} in our analysis. 

\appendix
\section{Best Practices for Achieving Q\&A Balance\footnote{Adapted from \url{https://github.com/jradavenport/Gender-in-Astro/wiki/Draft-of-Best-Practices}, a live version of this document.}}
\label{sec:prac}

Unlike a private interaction during a coffee break, the question and answer (Q\&A) interaction that follows a scientific talk is public, so there is the opportunity for everyone in the audience to learn from a question. There is also the opportunity for discussion, wherein another questioner comments on an exchange between a previous questioner and the speaker. Ultimately the goal is for everyone to learn something from the Q\&A interaction, not just the speaker and the questioner. In that case, it is important that we balance our Q\&A sessions to not only reflect the ideas and opinions for the most vocal and/or senior people, but to actively engage a broad range of individuals and ideas in our community.

This document focuses on gender and seniority because they are two of the most observable differences in our community and divide it into multiple large groups, but the ultimate goal is to have a conference and Q\&A session balance that reflects the underlying community and encourages contributions from marginalized groups. We welcome additional input on this topic.

\subsection{For Conference Organizers}
Studies indicate that Q\&A balance is related to SOC, speaker, chair and attendee balance \citep{Pritchard2014,Horner2016}. The first step for the conference organizers is to ensure balance when assembling in the SOC. In that case, it is useful to have the demographics of your community. The AAS is 25\% women and 73\% men \citep{Anderson2014}, while the gender balance of the International Astronomical Union (IAU) is recorded to be (as of 1 March 2017) 17\% women and 83\% men.\footnote{\url{https://www.iau.org/administration/membership/individual/distribution/}, retrieved 2 March 2017} Note that some subfields have a larger representation of women and an SOC drawn from those communities should have a larger representation of women.

There is significant material elsewhere on how to achieve speaker gender balance. One good example is ``Ten Simple Rules to Achieve Conference Speaker Gender Balance'' (\url{http://www.ncbi.nlm.nih.gov/pmc/articles/PMC4238945/}). Scheduling should also include significant time for question sessions, as the fraction of women asking questions increases with the number of questions allowed. If the conference includes independently organized splinter sessions, gender balance should be encouraged and enforced among the organizers, and they should be made aware of these or similar guidelines. 

\citet{Pritchard2014} recommends addressing equity directly in a plenary talk during the conference. We suggest having this early in a conference, perhaps even in the conference opening, so that more people are exposed to the talk. Certainly this would be a good time to also mention the conference's code of conduct and standards for behavior. 

\subsection{Session Chairs}
The biggest responsibility for Q\&A balance falls on session chairs because they shape the discussion. The session chairs should ideally be trained in person on some of the below guidelines and practices. If that is not possible, a list of guidelines should be supplied similar to the ones below.

\begin{itemize}
\item At the beginning of the session:
\begin{itemize}
\item The chair can specifically encourage junior scientists to ask questions and prepares senior people for not being called on immediately or at all. For example, ``There are many junior scientists in this session. I would like to encourage you to ask questions during the Q\&A’s.''
\item Set up expectations for keeping talks to time so the question session does not become shortened. 
\item Reinforce conference guidelines for accessibility (especially microphone use and leaving aisles clear).
\end{itemize}
\item After each talk: 
\begin{itemize}
\item Be strict on time. Use a timer and alarm, and move towards speakers who continue talking after their time is up. If a speaker goes over, penalize their Q\&A time, not other speakers.
\item Ask specifically for questions, rather than ``questions and comments.'' Comments often change the topic and tone of the conversation. 
\item Do not begin the Q\&A session with making a comment of your own, and reserve your question for the end (or ask in the absence of other questions). 
\item Ask for names and affiliations of individuals who ask questions (for recording purposes).
\item Do not immediately select the first hand to go up. Wait a few seconds to see if other hands go up, and take note of the balance. Priority should be given to women, junior scientists, and otherwise underrepresented individuals.
\item Aim for a total of 4--8 questions per talk, because at that point otherwise reticent questioners may be willing to raise their hands.
\end{itemize}
\item Moderating Discussion
\begin{itemize}
\item Do not allow questions to become speeches. Interrupt questioners who talk for a long time to request they shorten or end their time. Some example interventions include: ``Perhaps this is better for the coffee break.'' or ``Apologies, but for the sake of time can you ask your question?''
\item It is important to keep each Q\&A to time; do not continue to call on people when the time is up. 
\end{itemize}
\end{itemize}

\subsection{For Question Askers}
The people who ask questions at conferences can also contribute to a more fair and equitable conference question session. Some evidence indicates that if a women asks the first question, more women are likely to ask the subsequent questions \citep{Pritchard2014}. If you are a woman and/or an otherwise under-represented individual with a question, asking it can not only bring your ideas into the conversation but also encourage others to do the same. 

Conversely, (primarily male) participants who ask many questions should be conscious of the need to ``share the air'' with other attendees. Assuming the Cool Stars meeting talks with data were a representative set, we calculate that there were 332 total questions asked at CS18 and 487 at CS19. Given the number of attendees (364 and 477 respectively), that equates to 0.9 and 1 question asked per conference. Naturally, some people ask more and others none at all, but conference attendees who ask more than a few questions should be particularly conscious of asking questions that are of general interest and occasionally holding back to allow others time to participate. 

There will always be times when an audience member disagrees with the speaker about some scientific question. Even if that is the case, the best questions are both civil and concise and should be limited to information that could be generally useful to the community. One article on asking good questions\footnote{\url{https://www.theguardian.com/higher-education-network/2015/nov/11/dont-be-a-conference-troll-a-guide-to-asking-good-questions}} recommends that you should ``Register your dissent and the reasoning behind it without taking up too much time (or unleashing strong emotions). If your point is widely shared in the room, you won’t need to labour it; if it is not, a lengthy intervention will not win you many converts. You can always discuss your issues after the session.''

\setlength{\baselineskip}{0.6\baselineskip}
\bibliography{CSgender}
\setlength{\baselineskip}{1.667\baselineskip}

\end{document}